\documentclass[showpacs,twocolumn,amsmath,amssymb,pre,aps]{revtex4} 
\usepackage{graphicx}
\usepackage{bm}
\usepackage{mathrsfs}

\begin{document}
\title
{Effect of correlated noise on quasi-1D diffusion} 
\draft

\author{D.~V.~Tkachenko$^{1}$, V.~R.~Misko$^{1}$, and F.~M.~Peeters$^{1,2}$}
\affiliation
{$^{1}$Department of Physics, University of Antwerpen, Groenenborgerlaan 171,
B-2020 Antwerpen, Belgium}
\affiliation
{$^{2}$Departamento de F\'{\i}sica, Universidade Federal do
Cear\'{a}, 60455-900 Fortaleza, Cear\'{a}, Brazil}

\date{\today}

\begin{abstract}
Single-file diffusion (SFD) of an infinite one-dimensional chain of interacting
particles has a long-time mean-square displacement (MSD) $\propto t^{1/2}$, 
independent of the type of inter-particle repulsive interaction. 
This behavior is also observed in finite-size chains, although
only for certain intervals of time $t$ depending on the chain
length $L$, followed by the $\propto t$ for $t \to \infty$,
as we demonstrate for a closed circular chain of diffusing
interacting particles.
Here we show that spatial correlation of noise slows down
SFD and can result, depending on the amount of correlated
noise, in either subdiffusive behavior $\propto t^{\alpha}$,
where $0 < \alpha < 1/2$, or even in a total suppression of
diffusion (in the limit $N \to \infty$). 
Spatial correlation can explain the subdiffusive behavior in recent 
SFD experiments in circular channels. 
\end{abstract}
\maketitle


First introduced in bio-physics \cite{hodgkin} to account for
the transport of ions through molecular-sized channels
in membranes, the concept of single-file diffusion (SFD),
i.e., diffusion of particles in a narrow (quasi-1D) channel
where mutual passage is forbidden, has been extensively
studied during the last few decades
\cite{levitt,fedders,kaerger92}.

Recent advances in nanotechnology has stimulated a growing interest
in SFD, in particular, in the study of transport in nanopores.
Thus SFD was observed in experiments on diffusion of molecules
in zeolite molecular sieves \cite{meier}.
Zeolites with unconnected parallel channels serve as a
realization of the theoretically investigated one-dimensional systems.
Molecular diffusion of tetrafluoromethane in zeolite AlPO$_{4}$-5 was
studied by pulsed field gradient NMR spectroscopy \cite{hahn}.
The channel diameter was of the order of 0.73 nm, whereas the
diameter of the CF$_{4}$ molecules was 0.47 nm.
Thus mutual passage of molecules was excluded and it was found that 
the movement of an isolated particle, after a short ballistic period, 
was determined by the stochastic interaction with the channel walls 
yielding diffusional behavior, and the MSD was shown to obey an 
anomalous diffusion law. 

Despite the seemingly simplicity of SFD phenomenon, its theoretical
description involves considerable difficulties caused by the necessity
to solve the Fokker-Planck equation for the probability density
distribution in a multiparticle system.
Several studies tackled this problem for different model systems.
Thus in Ref.~\cite{PRB_28_5711} SFD of hard-core particles in a
one-dimensional lattice was theoretically investigated by the method
of ``tagged vacancies'' for an infinite chain, a finite chain with
reflecting boundaries and chains with periodic boundary conditions.
It was shown that for an infinite chain the mean-square
displacement (MSD) of a diffusing particle depends on time as
$\left\langle \Delta x^2\right\rangle\propto t^{1/2}$.
The same result was obtained in Ref.~\cite{kaerger92} and, it was
demonstrated in Ref.~\cite{Kollmann} that this result did not depend
on the nature of the interparticle interaction.
Note that this result was obtained assuming overdamped dynamics
of diffusing particles while the problem remains unsolved for the 
general case. 
For the intermediate case of neither overdamped, nor underdamped 
dynamics numerical simulations were presented in 
Ref.~\cite{weepl2007}. 

In case of a finite chain with reflecting boundaries, the long-time
MSD was shown \cite{PRB_28_5711} to become constant (i.e., independent
of time $t$) which means a total suppression of diffusion.
This result can be understood considering the fact that in a finite
system (single-connected, non-periodic and possessing no rotational
degree of freedom), no matter how large, only a finite set of
configurations of particles can be realized.
Thus for a long enough observation time, the system accesses all the
states in configuration space.
An exact analytical proof of this result using the Bethe-ansatz was
recently presented in Ref.~\cite{Lizana}.
A number of numerical studies on the SFD phenomenon have been performed
using Langevin dynamics (see, e.g., Ref.~\cite{Taloni}) including SFD
on periodic substrates \cite{MarchesoniPRL,MarchesoniPRE,Herrera}.

It is hard to fulfill the SF condition and investigate
SFD experimentally for atomic systems.
Alternatively, experimentalists used micro-meter sized colloidal
particles in narrow channels.
A well-defined SFD model was created \cite{Bechinger} by confining
paramagnetic
colloidal spheres of 3.6~$\mu$m in a set of circular trenches
(7~$\mu$m in width and 33 to 1608~$\mu$m in diameter)
fabricated by photolithography.
Using video microscopy, the trajectories of individual
particles were followed over long periods of time and revealed
a SFD behavior characterized by MSD $\propto t^{1/2}$.

SFD was recently realized even on a macroscopic scale.
The advantage of using a macroscopic system is that it allows one
to easily fulfill the SF condition and observe the motion of particles
using simple tools such as an optical microscope.
The diffusion of such macroscopic charged metallic balls
(of radius $R = 0.4$ mm and mass $m = 2.15$ mg)
was investigated in Ref.~\cite{coupier}.
The balls were interacting electrostatically and moving in a circular
channel such that mutual passing was forbidden, while mechanical
shaking induced an effective temperature.
The authors found that the system of interacting balls exhibited
a subdiffusive behavior $\propto t^{1/4}$, i.e., slower than that
predicted for SFD and observed in colloidal systems.

Here we investigate the effect of correlations of noise that causes 
diffusion. 
In particular, we study SFD of a finite-size system (i.e., interacting
particles diffusing in a closed circular channel) driven by a mixture
of a totally uncorrelated noise and a spatially-correlated (although
uncorrelated in time) noise. 
Being coordinate-independent, spatially-correlated noise can be also 
referred to as ``global''. 
We show that the presence of a spatially-correlated component dramatically
changes the SFD behavior, and it can also be used as an alternative
explanation of the slowing down of diffusion observed in experiments
with circular channels driven by artificial stochastic
noise modeling temperature.


{\it The model.---}
Let us consider SFD of interacting particles
in a circular channel driven, by non-correlated
{\it and spatially-correlated} stochastic forces.
The fact that both forces are stochastic implies the absence of time
correlations between the forces of the same component as well as
of mutual correlations between different components.
The fraction of the correlated noise to the total noise is given 
by a parameter $\lambda$, where $0 \leq \lambda \leq 1$. 
Therefore the fraction of non-correlated noise is $1-\lambda$.
The motion for the $i$-th particle is described by the Langevin equation,
\begin{eqnarray}
\label{eq1}
m \frac{d^2 \vec{r}_i}{dt^2} & = & -\gamma \frac{d\vec{r}_i}{dt} -
\sum_{j,i\neq j} \vec{\nabla U} (r_{ij}) -
\vec{\nabla U}_{conf}(r_{i}) \nonumber \\
& + & \lambda\vec{F}_{c}+ (1-\lambda)\vec{F}_{nc, i},
\end{eqnarray}
where $r_i$ is the radius of $i$-th particle, $m$ is its mass,
$\gamma$ is the friction coefficient.
The non-correlated part of noise
$\vec{F}_{nc, i}=\{F_{nc, i}^x(t), F_{nc, i}^y(t)\}$
and the spatially-correlated part
$\vec{F}_{c}=\{F_c^x(t), F_c^y(t)\}$
obey the following conditions,
\begin{eqnarray}
\label{eq4}
& & \langle F_{nc, i}^k(t^\prime)\rangle = 0,
\langle F_c^k(t^\prime)\rangle = 0, \nonumber \\
& & \langle F_{nc, i}^k(t^\prime)F_{nc, j}^l(t^{\prime\prime})\rangle =
2\gamma k_{B}T\delta_{i,j}\delta_{k,l}\delta(t^\prime-t^{\prime\prime}),
\nonumber \\
& & \langle F_c^k(t^\prime)F_c^l(t^{\prime\prime})\rangle =
2\gamma kT\delta_{k,l}\delta(t^\prime-t^{\prime\prime}),
\nonumber \\
& & \langle F_c^k(t^\prime)F_{nc, i}^l(t^{\prime\prime})\rangle = 0,
\end{eqnarray}
where $k,\ l={x,y}$ are the coordinate indices and $i,\ j$ are
the particle indices.
Such a choice of the correlation relations (\ref{eq4}) ensures that
the system eventually approaches an equilibrium state with temperature
$T$ at any time~\cite{MarchesoniPRL}.
The interparticle interaction potential $U(r_{ij})$ in our
calculations was taken a Yukawa potential,
$U(r)=\frac{q^2}{\varepsilon_0}\frac{e^{-\kappa r}}{r}$.
Here $\kappa = 1/\Lambda$ where $\Lambda = 4.8\cdot10^{-4}~m$
\cite{coupier}.
In our numerical calculations, other parameters of the system
were also taken from the experiment \cite{coupier}.
The general conclusions are independent of the exact functional 
form of the interaction potential. 
The confinement potential, $U_{conf}(r_i)$, which restricts the 
radial motion of the particles within the circular channel is 
taken parabolic: 
$U_{conf}(r_i)=\beta(r_{ch}-r_i)^2$,
where $\beta$ is chosen from the condition $V_{gs}=\beta r_0^2$,
where $r_0$ is the radius of the particle and $V_{gs}$ is the
ground state energy, i.e., the energy of the configuration when
all the particles are distributed equidistant
near the circular channel ``bottom''.


{\it Long-time SFD in a finite-size chain.---}
In Ref.~\cite{PRB_28_5711}, SFD of noninteracting (i.e., hard-core)
particles diffusing in a ring was studied theoretically using the
``tagged particle'' method.
It was shown that the long-time MSD grew as $2Dt$, where the
diffusion coefficient $D$ was inversely proportional
to the density of particles.

We will demonstrate that this result is also justified for
a closed (e.g., circular) chain of interacting particles.
For this purpose, let us first notice that SFD in a finite
system has two characteristic lengths: (i) the average
interparticle distance, $a$, and, as distinct from infinite
systems, (ii) the length of the chain, $L$.
The corresponding hierarchy of times raising from the
two characteristic lengths, implies that along with the 
``long-time'' regime when a particle diffuses for a distance
of the order of the average interparticle distance $a$
(during time $t_{a}$), there is another time scale, $t_{L}$,
which determines the long-time behavior.
This is the time required for a particle to diffuse over a
distance comparable to the system {\it length}, $L$.
Note that for large systems with $N \gg 1$ particles,
$t_{L} \gg t_{a}$ (and $t_{L} \to \infty$ when $N \to \infty$).
It is clear that due to the single-filing condition, diffusion
of {\it all} particles will be approximately the {\it same} for
$t \geq t_{L}$ and thus can be described by a ``collective''
coordinate $\vec{R} = (1/N) \sum_{i} \vec{r}_{i}$.

In case of a circular chain, one can define this collective
coordinate as an average angle of ``rotation'' of the system
as a whole,
$\phi = \langle \phi_{i} \rangle = (1/N) \sum_{i} \phi_{i}$.
The angular diffusion of $i$th particle is described by 
the system of equations which follow from Eqs.~(\ref{eq1}), 
\begin{eqnarray}
\label{eq10}
J_i \frac{\vec{d^2\phi_i}}{dt^2}
& = & -\gamma^{\prime}\frac{\vec{d\phi_i}}{dt}
- \sum_{j,i\neq j} \vec{r}_i\times\vec{\nabla U} (r_{ij})
\nonumber \\
& + & \lambda\vec{r}_i\times\vec{F}_{c}
+ (1-\lambda)\vec{r}_i\times\vec{F}_{nc, i},
\end{eqnarray}
where $J_i$ is the moment of inertia of $i$-th particle,
and $\gamma^{\prime} = \gamma r_{ch}^{2}$.
Taking the average over all the particles, we have:
\begin{eqnarray}
\label{eq11}
\frac{1}{N_p}\sum_i J_i\ddot{\phi}_i
& \simeq & J \ddot{\phi} = -\gamma^{\prime} \dot{\phi} + M_{inter}
\nonumber \\
& + & \lambda M_{c}+ (1-\lambda) M_{nc},
\end{eqnarray}
where we assume $J \simeq J_i$, i.e., neglect the radial dispersion,
and
\begin{equation}
\label{eq12}
M_{inter} = -\frac{1}{N_p}\sum_{i}\sum_{j,i\neq j}
\vec{r}_i\times\vec{\nabla} U(r_{ij}) = 0,
\end{equation}
\begin{eqnarray}
\label{eq13}
M_{c}
& = & \frac{1}{N_p} \sum_{i} \vec{r}_i \times \vec{F}_{c}
\nonumber \\
& \simeq & \oint_{channel} d \phi \vec{r}_i \times \vec{F}_{c} = 0,
\end{eqnarray}
\begin{eqnarray}
\label{eq14}
M_{nc}
& = & \frac{1}{N_p} \sum_{i} \vec{r}_i \times \vec{F}_{nc}.
\end{eqnarray}
It is worth noting that expression (\ref{eq12}) is exact due to
the sine theorem.
This well-known result from the dynamics of a conservative system
is applicable to the ``conservative part'' of our system
(although the whole system is not conservative). 
Relation (\ref{eq13}) means that the field of stochastic 
spatially-correlated force is homogeneous in space, so the integral 
of this force along any closed contour is equal to zero. 
In other words, the action of the spatially-correlated force on half
of particles in the channel is on average compensated by its action
on the other half of particles.
%
Thus the equation of motion for the average angle $\phi$ that 
characterizes the collective rotation of the system is: 
\begin{equation}
\label{eq16}
J \ddot{\phi} = -\gamma \dot{\phi} + (1-\lambda) M_{nc}.
\end{equation}
This equation is isomorphic to the equation of motion of a free 
particle driven by a stochastic force.
As a consequence, the long-time SFD behavior of a free particle can
be projected on the angular diffusion described by Eq.~(\ref{eq16}).
Integrating Eq.~(\ref{eq16}) we obtain the expression for the 
long-time angular MSD: 
$\left\langle \Delta \phi^2 \right\rangle = 2 D_{\phi} t$.
Thus we can conclude that long-time SFD (i.e., when $t \geq t_{L}$)
related to the collective rotation of the system of particles in a
circular channel is determined only by the stochastic
spatially-non-correlated force.
In other words, the interparticle interaction and stochastic
spatially-correlated force do not influence long-time diffusion.
On the other hand,
if the system is driven {\it only} by spatially-correlated noise
$(\lambda = 1)$, the stochastic term in Eq.~(\ref{eq16}) is zero, 
thus resulting in the total suppression of long-time SFD. 

We stress that these preliminary conclusions are based on 
Eqs.~(\ref{eq10})-(\ref{eq16}) for ``collective'' coordinate $\phi$ 
which are justified in the limit when $t \to \infty$ and $N \to \infty$ 
and can be used only for a qualitative prediction of the diffusive behavior 
in large ($N \gg 1$) finite systems in the long-time limit $t \geq t_{L}$. 
Our exact results presented below are obtained by numerically integrating 
Eqs.~(\ref{eq1}).

\begin{figure}
\begin{center}
\vspace*{0.5cm}
\hspace*{-0.5cm}
\includegraphics*[width=7.0cm]{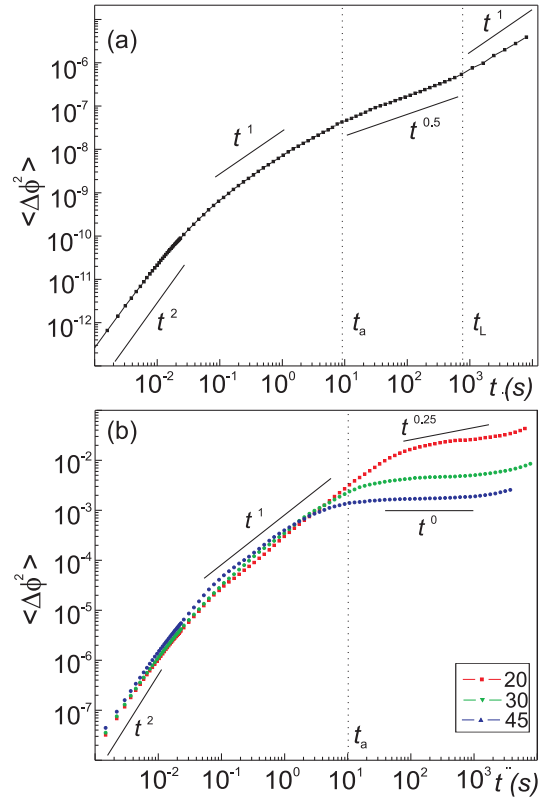}
\end{center}
\vspace{-0.5cm}
\caption{
(Color online)
The average MSD (in log-log scale) of a particle diffusing
in a circular channel with $N$ particles, driven by:
(a) purely stochastic noise ($\lambda = 0$), for $N = 20$, and
(b) fully spatially-correlated noise ($\lambda = 1$), for $N = 20$,
30, and 45.
$\phi$ is measured in radians (also in Fig.~2).
For $\lambda = 0$, the initial fast grow $\propto t^{2}$ and
$\propto t^{1}$ of the MSD is followed by the
$\propto t^{1/2}$-region (for $t_{a} \leq t \leq t_{L}$) which gradually evolves
to the long time ($t > t_{L}$)
$\propto t^{1}$ asymptotic behavior (a).
For $\lambda = 1$ (b), the $t^{1}$ growth rapidly changes
by a subdiffusive behavior $\propto t^{\alpha}$, with $0 < \alpha < 1/2$.
}
\vspace{-0.5cm}
\label{diffusion}
\end{figure}

The results of numerical calculations of the MSD of the SFD of $N$
particles in a circular channel are shown in Fig.~1 for two limiting
cases:
(i) for $N = 20$, $\lambda = 0$, i.e., for purely non-correlated noise
(Fig.~1(a)),
and
(ii) for $N = 20$, 30, and 45, $\lambda = 1$, i.e., for fully
spatially-correlated noise (Fig.~1(b)).
In case $\lambda = 0$, the initial fast grow (i.e., $\propto t^{2}$
followed by $\propto t^{1}$) of the MSD is followed by the pronounced
$\propto t^{1/2}$ ``long-time'' (for infinite systems) region
(for $t_{a} \leq t \leq t_{L}$) gradually changing to 
$\propto t^{1}$ asymptotic behavior for $t > t_{L}$ 
(Fig.~1(a)) 
which is common for any finite chain and any $\lambda$. 
The short-time behavior for $\lambda = 1$ is rather similar,
while for longer time the $t^{1}$ growth rapidly changes to
a subdiffusive behavior
$\propto t^{\alpha}$, with $0 < \alpha < 1/2$,
gradually approaching a $t^{0}$-plateau for large $N$ 
\cite{largen}
(see Fig.~1(b)), in agreement with Eq.~(\ref{eq16}).


{\it Spatially-correlated vs. non-correlated noise.---}
In our numerical study of angular diffusion in a circular channel
the standard deviation of an average angle (i.e., the MSD),
\begin{equation}
\label{eq18}
\left\langle\left\langle \Delta\phi^2\right\rangle_{e}\right\rangle_{p}=
\frac{1}{N_{p}N_{e}}\sum_{i,j}
\left( \left\langle \Delta \phi_{i,j}^2\right\rangle_t
- \left\langle \Delta \phi_{i,j}\right\rangle_t^2 \right)
\end{equation}
was calculated as a function of time, $t$.
Here $\left\langle...\right\rangle_{p}$ denotes averaging over all
particles of the given ensemble,
$\left\langle...\right\rangle_{e}$ over various ensembles,
and
$\left\langle...\right\rangle_{t}$ over time.
The number of ensembles was chosen typically from $100$ to $300$,
for $N=5$ to 45 particles in the ensemble.

\begin{figure}
\begin{center}
\includegraphics*[width=7.0cm]{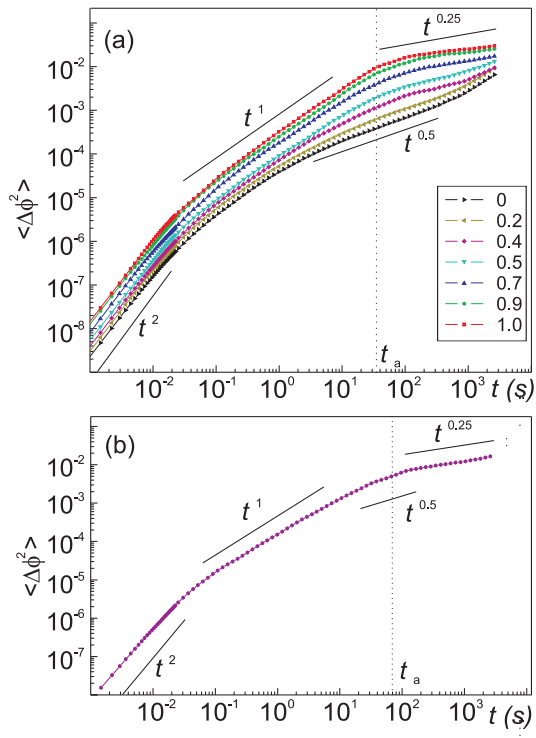}
\end{center}
\vspace{-0.5cm}
\caption{
(Color online)
The average MSD (in log-log scale) of a particle, driven by
a mixture of non-correlated and spatially-correlated noise,
for:
$\lambda = 0$, 0.2, 0.4, 0.5, 0.7, 0.9, and 1 (a).
With increasing $\lambda$,
the region of fast growth $\propto t$ expands,
while the region of subdiffusive growth, i.e., $\propto t^{1/2}$,
shrinks and rapidly changes to a stronger subdiffusive mode
$\propto t^{\alpha}$, with $0 < \alpha < 1/2$, including
$\propto t^{1/4}$ for $\lambda = 0.6$ (b).
}
\vspace{-0.5cm}
\label{diffusion}
\end{figure}

Results of simulations for different $\lambda$ show (Fig.~2(a))
that when the fraction of the correlated noise increases,
the region of fast growth (similar to the behavior of a free
particle) of SFD $\propto t$ expands,
while the region of subdiffusive growth, i.e., $\propto t^{1/2}$
(typical for SFD of a particle in an infinite chain),
is essentially narrowed and rapidly changes to even stronger
subdiffusive mode
$\propto t^{\alpha}$, with $0 < \alpha < 1/2$.
Thus we observe a considerable slowing down of diffusion on a long-time 
scale when increasing the fraction of spatially-correlated noise. 
Note that for some value of $\lambda$, i.e., $\lambda \approx 0.6$,
the long-time $\alpha$ becomes $\approx 0.25$, i.e., leading to
a pronounced region in the MSD $\propto t^{1/4}$ (Fig.~2(b)).
In the limit of completely spatially-correlated noise ($\lambda = 1$),
the calculated MSD
$\left\langle \Delta \phi^2\right\rangle$
saturates for large time $t$
(for sufficiently large number of particles, see Fig.~1(b)).

The observed saturation of the MSD for long-time scale looks similar
to that found in a finite chain where the motion of particles was
restricted in the direction of the chain that in turn was related
to the fact that for long time (of the order of the time of diffusion
of a particle over the whole system) the trajectory of the particle
filled practically all the configurational space accessible to the
given particle.
However, in spite of the similarity in the long-time SFD behavior
with, e.g., a finite chain of particles in the presence of reflecting
boundaries \cite{Lizana}, 
the physics behind the observed slowing down of the diffusion 
in our system is completely different. 
Our system is periodic and it possesses a rotational degree of freedom
which leads to a single-particle-like long-time asymptotic SFD behavior
(as shown in Fig.~1(a)).
Adding a fraction of spatially-correlated noise effectively suppresses
the rotational degree of freedom and thus the SFD in the long-time
limit.

Our results provide an alternative explanation of the subdiffusive
behavior (i.e., slower than $\propto t^{1/2}$) of particles diffusing
in a circular channel driven by an {\it artificial} stochastic force.
For example, in SFD experiments with charged balls \cite{coupier},
the system temperature was modeled by random acoustic waves produced
by a set of loudspeakers placed under the substrate with a circular
channel.
Thus the {\it whole} sample was shaked and, consequently, different
particles were {\it not} driven individually.
The spatial correlation of noise (although being uncorrelated in time)
in this system arises, in our opinion, due to two factors:
(i) the collective excitation of all the particles in the system by
a limited set of sources of noise (note that even a large number of
different sources would lead to spatial correlation of noise, due to
the interference between the waves), and
(ii) shaking the whole sample containing the channel with all the
particles embedded in a single-file chain.
Note that as was suggested by the authors of Ref.~\cite{coupier}
and confirmed by recent numerical simulations \cite{weepl2007},
the interparticle interaction in a non-overdamped regime perhaps
could also lead to a slowing down of the long-time SFD in circular
channels.

%
Similar conditions, leading to spatially-correlated noise, can also be
realized in other artificial systems that exhibit single-file behavior.


{\it Conclusion.--- }
We have investigated the effect of spatial correlation in noise 
on single-file diffusion of interacting particles diffusing in a 
finite-size circular channel 
where the long-time diffusion is characterized by two time regimes, 
$t_{a} < t < t_{L}$ and $t > t_{L}$. 
We demonstrated that in the limit of purely non-correlated noise,
the mean-square displacement characterizing the diffusion rate 
shows a $\propto t^{1/2}$ (for $t_{a} < t < t_{L}$) and 
a $\propto t^1$ (for $t > t_{L}$) behavior independent of the 
type of the interparticle interaction. 
Adding a fraction of spatially-correlated noise slows down the
diffusion and could result in either subdiffusive behavior
$\propto t^\alpha$ with $\alpha < 1/2$ (for $t_{a} < t < t_{L}$), 
or even in the total suppression of diffusion on a long-time scale 
(for $N \to \infty$). 
Our model provides an alternative explanation of experimentally 
observed slowing down of the diffusion rate of interacting particles 
driven by artificial noise sources.

\smallskip

We acknowledge discussions with M.~Saint-Jean.
This work was supported by the ``Odysseus'' program of the
Flemish Government and
FWO-Vl.


\begin{thebibliography}{0}
\bibitem{largen} 
Note that the $t^{0}$ regime is achieved only 
for $N \to \infty$ when $t_{L} \to \infty$ 
and thus the long-time regime $t \geq t_{L}$ 
cannot be reached.
\end{thebibliography}
\end{document}